\documentstyle[12pt,emulateapj]{article}

\begin{document}

\lefthead{S. Nitta}
\righthead{Reconnection rate in the self-similar evolution model}

\title{Magnetic Reynolds number dependence of reconnection rate and flow structure of the self-similar evolution model of fast magnetic reconnection}

\author{Shin-ya Nitta}
\affil{Hayama Center for Advanced Studies, 
The Graduate University for Advanced Studies (SOKENDAI), 
Shonan Village, Hayama, Kanagawa, 240-0193 Japan; snitta@koryuw02.soken.ac.jp}
\authoremail{snitta@koryuw02.soken.ac.jp}

\begin{abstract}

This paper investigates Magnetic Reynolds number dependence of the ``self-similar evolution model'' (Nitta et al. 2001) of fast magnetic reconnection. I focused my attention on the flow structure inside and around the reconnection outflow, which is essential to determine the entire reconnection system (Nitta et al. 2002). The outflow is consist of several regions divided by discontinuities, e.g., shocks, and it can be treated by a shock-tube approximation (Nitta 2004). By solving the junction conditions (e.g., Rankine-Hugoniot condition), the structure of the reconnection outflow is obtained. Magnetic reconnection in most astrophysical problems is characterized by a huge dynamic range of its expansion ($\sim 10^7$ for typical solar flares) in a free space which is free from any influence of external circumstances. Such evolution results in a spontaneous self-similar expansion which is controlled by two intrinsic parameters: the plasma-$\beta$ and the magnetic Reynolds number. The plasma-$\beta$ dependence had been investigated in our previous paper. This paper newly clarifies the relation between the reconnection rate and the inflow structure just outside the Petschek-like slow shock: As the magnetic Reynolds number increases, strongly converging inflow toward the Petschek-like slow shock forms, and it significantly reduces the reconnection rate. 

\end{abstract}


\section{Introduction}
\label{sec:Int}
Magnetic reconnection is one of probable candidates for very powerful energy conversion mechanism in magnetized plasma systems, e.g., solar flares and geomagnetospheric substorms. Majority believes that magnetic reconnection is a universal mechanism also for recent energetic topics, e.g., accretion disk X-ray emission of young stellar objects (YSOs; see  Koyama et al. 1994; Hayashi et al. 1999) and galactic ridge X-ray emissions (GRXEs; see Koyama et al. 1986; Tanuma et al. 1999). 

There are so many phenomenological studies of magnetic reconnection mainly by numerical simulation researchers, however, essential open questions still remain, not only regarding the microscopic physics of the anomalous resistivity, but also the macroscopic magnetohydrodynamical (MHD) structure. Especially, understanding of the coupling process between different scales, i.e., how does the microscopic scale process relating to the origin of anomalous resistivity ($\sim 10^0$[m] for solar corona) govern the macroscopic scale process in which enormous magnetic energy releases ($\sim 10^7$[m] for typical flare loop) is very important, but still an open question. One may think it is very curious that extremely small diffusion region governs the huge entire reconnection system (e.g., idea of the ``fractal diffusion region''; see Shibata \& Tanuma 2001). The aim of the series of our papers is to clarify the relationship and causality between these two different scales in a regime of MHD (see subsection \ref{sec:scale-coupling}). In this work, our attention is focused on the magnetic Reynolds number dependence of the flow structure and the reconnection rate. 

We consider a magnetic energy conversion in an antiparallel magnetic equilibrium configuration called a ``current sheet system.'' In such the case, the electric resistivity plays an important role: It breaks the frozen-in and leads to the magnetic diffusion. The magnetic diffusion can release the magnetic energy by the Ohmic dissipation. However since the Spitzer resistivity of the most cases of astrophysical plasma systems is considerably small, the magnetic diffusion speed should be very small. This means that the magnetic diffusion itself is not efficient for a quick energy conversion like in the solar flares. We here introduce an essential non-dimensional parameter so-called the ``magnetic Reynolds number $R_{em}$'' which is defined as the ratio of the Alfv\'{e}n wave transit timescale across the system to the magnetic diffusion timescale. The most of astrophysical plasma is highly conductive  (e.g., $R_{em} \sim 10^{14}$ for typical solar corona), thus the energy conversion speed by the Ohmic dissipation is very slow (the energy conversion timescale is of the order of $10^8$[yr]!!: This is too long to explain the solar flares). 

It is important that, even in such the case, the processes involving magnetic reconnection can release the magnetic energy at a very large power. The mostly successful reconnection model for astrophysical applications is the Petschek model (see Petschek 1964). Although the reconnection of magnetic field lines itself is caused by the non-uniform magnetic diffusion, the main energy conversion is owing to the propagation of the X-shaped slow shocks in the Petschek model. Consequently, the Petschek model can convert the magnetic energy in a very short time (the wave propagation timescale estimated as of the order of $10^2$[sec] for typical solar flares) almost independent of the magnetic Reynolds number. We call such a very powerful process in which the magnetic diffusion timescale does not determine the entire energy conversion as the fast reconnection. As discussed above, the Petschek model might be the most significant model to explain actual reconnection phenomena, but the author thinks there is an essential lack to be a real astrophysical reconnection model. The following part of the introduction is focused on this point.

We must note that many cases of the actual magnetic reconnection in astrophysical systems usually grow over a huge dynamic range in its spatial dimension. For example, the initial scale of the reconnection system can be defined by the initial current sheet thickness, but this is too small to be observed in typical solar flares. We do not have any convincing estimate of the scale, but if we estimate it to be of the order of the ion Larmor radius, it is extremely small ($\sim 10^0$ [m] in the solar corona). Finally, the reconnection system develops to a scale of the order of the initial curvature radius of the magnetic field lines ($\sim 10^7$ [m] $\sim$ 1.5\% of the solar radius for typical solar flares). The dynamic range of the spatial scale is obviously huge ($\sim 10^7$ for solar flares). For geomagnetospheric substorms, their dynamic range of growth is also large ($\sim 10^4$ for substorms). Such a very wide dynamic range of growth suggests that the evolution of the magnetic reconnection should be treated as a development in ``free space'', and that the external circumstance does not affect the evolutionary process of a magnetic reconnection, at least at the expanding stage just after the onset of reconnection. 

Even a system is not completely free from the influence of the external circumstances, we can approximately treat it as a spontaneously evolving system if the evolution timescale (which is estimated as the Alfv\'{e}n transit time for the final system scale because the spontaneous expansion speed of the reconnection system is equivalent to the fast-mode propagation speed, see Nitta et al. 2001; hereafter paper 1) is much smaller than the timescale imposed by the external circumstances (e.g., the convection timescale). For typical substorms, the evolution timescale of the reconnection is of the order of second or minute while the convection timescale to compress the current sheet is of the order of hour. In such cases, the external circumstances simply play a role of triggering the onset of a reconnection and the evolution itself is approximately free from the influence of the external circumstances. Though, of course, the exceptional cases in which the external circumstances intrinsically influence the evolution can arise for the particular situation (e.g., the cases that very fast convection drives the reconnection), the author believes that it is worth to establish a new model applicable to the cases which are free from the external influences. 

However, no reconnection model evolving in a free space has been studied. We should note that the most of previous theoretical (e.g., Petschek 1964, Vasyliunas 1975, Priest \& Forbes 1986) and numerical works on a reconnection treated it as a boundary problem strongly influenced by external circumstances. Our interest is focused on a spontaneous evolutionary process of a magnetic reconnection in a free space as a macroscopic instability in a current sheet system. This evolutionary reconnection model can explain a coupling process between different scales, i.e., a microscopic scale represented by the ion Larmor radius and a macroscopic scale represented by the scale of external circumstances. 

The author and collaborators had been studied such the problem of two-dimensional MHD reconnection in a series of papers. A non-linear evolution of the spontaneous reconnection in a free space is numerically simulated (two-dimensional MHD simulations with artificially enhanced localized resistivity) in paper 1. The result of our numerical simulation clearly shows a self-similar expansion of the reconnection system. Owing to technical reasons (restriction of CPU time, memory etc.), we could confirm the stable expansion of the reconnection system only in a dynamic range of $10^2-10^3$ in spacial scale of the expansion. This dynamic range is obviously insufficient because actual dynamic range seems to be $10^4-10^7$ as discussed above. In order to supplement the lack of the dynamic range, we studied the inflow region of the reconnection system semianalytically by so-called the ``Grad-Shafranov approach'' (Nitta et al. 2002; hereafter paper 2) with a boundary condition along the Petschek-like X-shaped slow shock (Nitta 2004; hereafter paper 3). We obtained a self-similarly expanding solution which is fairly consistent with our numerical result in paper 1. Therefore we can conclude that the self-similar solution of an expanding reconnection system is stable with any ideal MHD modes and will continue to expand until the system scale becomes to be comparable to the typical scale of the external environment, e.g., the diameter of a magnetic flux loop. In paper 3, we studied a structure of the reconnection outflow which provides the boundary condition along the Petschek-like slow shock by a shock-tube approximation. This boundary condition is very important because it determines the structure of the inflow region (see paper 2). 

The spontaneously expanding reconnection system has only two parameters: the plasma-$\beta$ value and the magnetic Reynolds number. We could clarify the plasma-$\beta$ dependence for a very wide dynamic range of plasma-$\beta$ in paper 3 (a brief summary is in the next section). This enables us to apply our self-similar reconnection model to a wide variation of phenomena which may have different plasma-$\beta$. In this paper, we study the magnetic Reynolds number dependence of the reconnection system. While majority believes that a spatially localized anomalous resistivity is important to realize a fast reconnection (e.g., Yokoyama \& Shibata 1994), we do not have any established model for the mechanism and a convincing estimated value of the anomalous resistivity (Coppi \& Friedland 1971, Ji et al. 1998, Shinohara et al. 2001) yet. Thus, it is important to study the magnetic Reynolds number dependence of the properties of the reconnection system. We have obtained the following results. 

We define the effective magnetic Reynolds number as $R_{em}^* \equiv V_{A0}/V_{dif}^*$ where $V_{A0}$ is the Alfv\'{e}n speed at the asymptotic region (the region far from the current sheet) and $V_{dif}^*$ is the actual magnetic diffusion speed at the diffusion region.\\

\noindent
1) Strongly converging inflow\\
As $R_{em}^*$ increases, the convergence of the inflow toward the Petschek-like slow shock@becomes significant. This leads that the inflow velocity and the magnetic field line at the inflow region tend to be parallel to each other.\\

We here define the reconnection rate $R^*$ as the non-dimensional reconnection electric filed (the ratio to the product of $V_{A0}$ and $B_0$ [the magnetic field strength at the asymptotic uniform region]). \\

\noindent
2) Reconnection rate\\
As a result of the converging inflow, the reconnection electric filed decreases as $R_{em}^*$ increases. Finally we find $R^* \propto 1/R_{em}^*$ at the large magnetic Reynolds number limit.  

This paper is organized as follows. We state a model of the reconnection outflow as a kind of shock-tube problem in section \ref{sec:scheme}. The basic equations for the outflow structure and the numerical procedure to solve the equations are listed also in this section. By solving the basic equations, we obtain the flow structure also the reconnection rate. The results are shown in section \ref{sec:result}. We discuss the properties of these results in section \ref{sec:discussion}.

\section{Numerical Scheme}
\label{sec:scheme}

\subsection{Concept}
\label{sec:concept}

As we mentioned in paper 3, the reconnection outflow in our spontaneous model has a structure involving several discontinuities like a shock-tube problem. We solve the reconnection outflow region by approximating the problem as a kind of shock-tube problem (see details in the next subsection). In paper 3, we assumed a non-converging inflow toward the slow shock approximating the original Petschek model, and discussed the structure of the outflow region as a function of the ambient plasma-$\beta$ value. Consequently, we obtain the plasma-$\beta$ dependence of the spontaneous inhalation speed (see figure \ref{fig:sp-in}) and the reconnection rate (see figure \ref{fig:R-beta0}). These figures are essentially the same with the corespondent figures in paper 3, but newly plotted by a more precise calculation.  

In this paper we are willing to study the magnetic Reynolds number dependence of the outflow structure by the above shock-tube approximation in the regime of ideal (non-resistive) MHD. One may think that it is very curious because we cannot treat the magnetic Reynolds number in the regime of ideal MHD as in this work. We should note, however, that if we can evaluate the reconnection electric field $E_z$ (note that $E_z$ is uniform in the vicinity of the reconnection point from the Ampere's law because this region is almost stationary in our self-similar model) and the magnetic field $B_x^*$ around the diffusion region from the shock-tube problem, we can estimate the inflow speed $v_y^*\ (\equiv E_z/B_x^*)$ toward the diffusion region. This inflow speed is in balance with the magnetic diffusion speed $v_{dif}^*$ at the diffusion region. Providing that the Alfv\'{e}n speed $V_{A0}$ at the asymptotic region to be fixed as a normalization value as in paper 3, inverse of the inflow speed toward the diffusion region shows the effective magnetic Reynolds number $R_{em}^*$ ($R_{em}^* \equiv V_{A0}/v_{dif}^*=V_{A0}/v_y^*$). Thus we can treat the magnetic Reynolds number in a context of ideal MHD. We would stress again that our attention is focused on the magnetic Reynolds number dependence of the reconnection system, i.e., the outflow structure and the reconnection rate.

\subsection{Model}
\label{sec:model}

We present a schematic picture of the reconnection outflow in our self-similar evolution model (see figure \ref{fig:sch}). The coordinates are defined as follows: $x$-axis is parallel to both the current sheet and initial anti-parallel magnetic field, $y$-axis is perpendicular to the current sheet, and $z$-axis is parallel to the current sheet, but is perpendicular to the initial magnetic field (hence, $\partial_z$=0 in this two-dimensional problem). The essence of the model is quite the same as the model in paper 3. A similar structure of the reconnection outflow is shown in figure 9 of Ugai (1999) as a simulation result. Because of symmetry with respect to $x-$ and $y-$axes, we treat only the region $x,\ y>0$. The outflow is composed of two different plasmas which have different origins. These two plasmas are touch in contact at the point $x=x_c$ (the contact discontinuity). A plasmoid forms around the contact discontinuity. The rear-half region ($x < x_c$: the reconnection jet) is filled with the reconnected plasma coming from outside the current sheet (the inflow region). The front-half region ($x > x_c$: the plasmoid) is filled with the original current sheet plasma. 

The entire outflow is surrounded by a slow shock which has a complicated `crab-hand' shape (see Abe \& Hoshino 2001). The Petschek-like figure-X shaped slow shock (an oblique shock) is elongated from the diffusion region with a slight opening angle $\theta$. There is a reverse fast shock (an almost perpendicular shock) inside the reconnection jet ($x=x_f$). In front of the plasmoid, a forward V-shaped slow shock (an oblique shock) forms. The opening angle and the locus (the crossing point with $x-$axis) are $\phi$ and $x_s$, respectively. 

The entire structure including the several discontinuities is analogous to the one dimensional shock-tube problem. The forward shock (a V-shaped slow shock) and the reverse shock (a fast shock) are formed by the collision of the reconnection jet and the original current sheet plasma, and propagate in both directions. Between these two shocks, the contact discontinuity forms. We approximate this reconnection outflow as a quasi one-dimensional problem in order to solve it analytically. Such an approximation may be valid near the $x-$axis, because the system is symmetric with respect to the $x-$axis. 

We focus our attention on the quasi-one-dimensional problem along the inflow stream line and the reconnection outflow. We treat the figure L-shaped region $x, y =finite \gg D$ if the inflow stream line is parallel to the  $y-$axis where $D$ is the initial current sheet thickness. Note that this region apparently tends to coincide with the $x-$ and $y-$axes in the self-similar stage which is a very late stage from the onset when we observe the evolution in a zoom-out coordinate [see paper 1]. If the inflow stream lines are inclined owing to the convergence of the inflow, we consider the modified L-shaped region along the inflow stream line (see SSL of figure \ref{fig:channel}) in stead of the $y-$axis. Each region between two neighboring discontinuities is approximated to be uniform. We note about the up-stream region p just above the X-shaped slow shock that the region between the slow shock and the separatrix field line (X-shaped field line reaching the X-point) is also approximated to be uniform. In this region, each reconnected field line has an almost straight shape and crosses the X-slow shock while each field line has hyperbolic shape in the region above the separatrix field line. In paper 3, we presumably suppose a non-converging inflow in the region p for the spontaneous inhalation, but if inhalation speed is reduced when $R_{em}^*$ increases, it may result in an alteration of the inflow structure. Hence, we should take a possibility of convergence or divergence of the inflow in the region p into our problem (see below and subsection \ref{sec:procedure}).  

The region 3 between the contact discontinuity and the V-shaped slow shock looks to be non uniform. Since we do not know how we can solve the two dimensional structure of this region analytically, we approximate this region to be uniform. 

The quantities denoting the initial uniform equilibrium at the asymptotic region are gas pressure $P_0$, mass density $\rho_0$ and magnetic field strength $B_0$. The plasma-$\beta$ value at the asymptotic region is defined as $\beta_0 \equiv P_0/(B_0^2/[2 \mu])$ where $\mu$ is the magnetic permeability of vacuum. In the rest of this paper, we use the normalization of physical quantities as like in paper 2. We define units for each dimension as follows: (unit of velocity)$=V_{A0} \equiv B_0/\sqrt{\mu \rho_0}$ (Alfv\'{e}n speed at the asymptotic region), (unit of the length)$=V_{A0} t$ where $t$ is the time from the onset of reconnection, (unit of mass density)$=\rho_0$, (unit of magnetic field)$=B_0$, (unit of pressure)$=\beta_0/2 \cdot \rho_0 V_{A0}^2$.  

We set the quasi-one-dimensional shock-tube-like problem as follows. The system has 22 unknown quantities: $P_p$, $\rho_p$, $v_{xp}$, $B_{xp}$, $B_{yp}$, $\theta$, $P_1$, $\rho_1$, $v_1$, $B_1$, $x_f$, $P_2$, $\rho_2$, $v_2$, $B_2$, $P_3$, $\rho_3$, $v_{y3}$, $B_{x3}$, $B_{y3}$, $\phi$ and $x_s$ where $P_*$, $\rho_*$, $v_*$, $B_*$ denote the pressure, density, velocity, magnetic field, respectively (Note $x_c=v_{x3}=v_2$ because no mass flux passes through the contact discontinuity). The suffices $p$, 1, 2 or 3 denote the region divided by the discontinuities. The suffices $x$ or $y$ denote the vector components. Quantities $\theta$, $x_f$, $\phi$ and $x_s$ denote the inclinations of the X-shaped slow shock, the locus of the fast shock, the inclination of the V-shaped slow shock and the locus of the V-shaped slow shock (crossing point with $x-$axis), respectively. 

These unknowns should be related to each other via conditions coming from the integrated form of conservation laws (i.e., the Rankine-Hugoniot [R-H] conditions) or other relations. The set of relations is listed in the next subsection. 

There is an essential difference from the model in paper 3. In paper 3, we assumed (or imposed) the transversal component of the inflow velocity vanishes ($v_{xp}=0$) and solved the vertical component $v_{yp}$ of the inflow velocity (and other unknowns). We here treat $v_{yp}$ as an artificially controllable variable and its value is given by hand. The change of $v_{yp}$ may induce a change of the strength of the fast-mode rarefaction. Since the inflow is induced by the fast-mode rarefaction, this change may result in a change of $v_{xp}$. Hence, we must solve $v_{xp}$ as a new unknown quantity.

\subsection{Basic equations}
\label{sec:b-eqs}
There is no essential difference of the basic equations for the outflow structure between the equations used in paper 3 except that we must prepare the equations denoting the oblique inflow case $v_{xp} \neq 0$. In the most cases, we must simply add several terms involving $v_{xp}$ (e.g., in equations 3, 5, 6, 7, 13 and 22 in paper 3). 

In equation 1 (the frozen-in condition at the region p), a physical insight to modify it is required. In the oblique ($v_{xp} \neq 0$) inflow case, we should note that the frozen-in condition requires that the ratio of the perpendicular component $B_\perp$ of the magnetic field with respect to the inflow velocity versus the mass density holds to be a conservative quantity. If we can presumably obtain the stream line configuration all over the inflow region, we can state exact modified equation for the frozen-in condition, however, we never know it in the context of this quasi-one-dimensional shock-tube approximation. Hence we need an approximating model of the inflow stream line configuration: we here simply approximate it to be straight. Thus we obtain equation \ref{eq:frozen-in} in appendix. 

According to the above model of the reconnection outflow, we obtain the following 22 equations for 22 unknown quantities (Detailed forms of each equation are listed in the appendix). \\

\noindent
1)relations between pre-X-slow shock and asymptotic region \\
Frozen-in condition [eq1]\\
Polytropic relation [eq2]\\

\noindent
2)R-H jump conditions at X-slow shock\\
Pressure jump [eq3]\\
Density jump [eq4]\\
Velocity jump (parallel comp.) [eq5]\\
Velocity jump (perpendicular comp.) [eq6]\\
Magnetic field jump (parallel comp.) [eq7]\\
Magnetic field jump (perpendicular comp.) [eq8]\\

\noindent
3)R-H jump conditions at reverse fast shock\\
Pressure jump [eq9]\\
Density jump [eq10]\\
Velocity jump [eq11]\\
Magnetic field jump [eq12]\\

\noindent
4)Magnetic flux conservation at X-point [eq13]\\

\noindent
5)Force balance at contact discontinuity [eq14]\\

\noindent
6)R-H jump conditions at forward V-slow shock\\
Pressure jump [eq15]\\
Density jump [eq16]\\
Velocity jump (parallel comp.) [eq17]\\
Velocity jump (perpendicular comp.) [eq18]\\
Magnetic field jump (parallel comp.) [eq19]\\
Magnetic field jump (perpendicular comp.) [eq20]\\

\noindent
7)Boundary Condition at the tip of the outflow [eq21]\\

\noindent
8)Magnetic flux conservation all over the outflow [eq22]\\

We can solve the majority of these equations by hand, and by substituting the solutions into other equations, we can reduce equations. Finally, ten equations 1, 3, 4, 5, 6, 7, 11, 14, 21 and 22 remain as complicated nonlinear coupled equations for ten unknowns $\rho_p$, $v_{xp}$, $B_{xp}$, $B_{yp}$, $\theta$, $P_1$, $\rho_1$, $x_f$, $\phi$ and $x_s$.

\subsection{Numerical procedure}
\label{sec:procedure}
We solve these coupled ten equations by an iterative method (the Newton-Raphson method). In order to find the well converging solution of the nonlinear coupled equations, a precise initial guess of the unknowns is required. In general, this step to find an appropriate initial guess is the core of difficulty (this should be comparable to a ``treasure hunting in ten-dimensional space'' with an incomplete treasure map). Fortunately, this most difficult step had already been cleared in paper 3 for the case $v_{xp}=0$. 

We can start from the solution of the case $v_{xp}=0$ for an arbitrary value of $\beta_0$. Our most interest is focused on low-$\beta$ cases typically appear in astrophysical applications. We demonstrate the numerical procedure for the case $\beta_0=0.01$ as the reference case because this is a typical value in solar corona and geomagnetosphere. In the work of paper 3, we have already had the solution in the spontaneous inhalation case ($v_{xp}=0$) for $\beta_0=0.01$.

First, we start with the solution of the case $v_{yp} = -v_{insp}(0.01) = -0.06398$ (where $v_{insp}(0.01)$ is the spontaneous inhalation speed for $\beta_0=0.01$) as the initial guess for the case $v_{yp}=-v_{insp}(0.01)+\Delta v_{y}$, where $\Delta v_{y}$ is the increment of $v_{yp}$. Once we find a converged solution of the Newton-Raphson procedure for the case $v_{yp}=-v_{insp}(0.01)+\Delta v_{y}$, we treat it as the initial guess for the case $v_{yp}=-v_{insp}(0.01)+2 \Delta v_{y}$, and we have successively obtained the solutions for different $v_{yp}$. Of course, $\Delta v_{y}$ should be a small enough value to keep good convergence of the Newton-Raphson method. From several trials, we carefully adopt the following two cases of the increment: $\Delta v_{y}=10^{-6}$ and $10^{-7}$. We have checked that the results of these two cases fairly coincide with each other. 

We obtain the result that, as $|v_{yp}|$ decreases, the strength of the reverse fast shock decreases. At a critical value of $|v_{yp}|$, the reverse fast shock no longer forms (the density jump ratio and the pressure jump ratio reduce to unity). At this point we exchange the numerical code to another one for the case of no reverse fast shock forms (see the last part of appendix), and we can continue the calculation. Above procedure is valid for any value of $\beta_0$.

\section{Result}
\label{sec:result}

We have investigated the response of the reconnection outflow structure to a variation of $v_{yp}$ (y-component of the inflow velocity at the pre-slow shock region [region p] normalized by the ambient Alfv\'{e}n speed $V_{A0}$). Usually, it shows the reconnection rate if the transverse component $B_{xp}$ of the magnetic field is almost same as its ambient value $B_0$ and x-component $v_{xp}$ of the inflow velocity is negligible ($|v_{xp}| \ll |v_{yp}|$). We must note, however, that if a significant converging inflow arises or the transverse magnetic component reduces from its ambient value, we need an alternative estimation of the reconnection rate. Actuary a significant converging inflow and decrease of the transverse magnetic component realize when $|v_{yp}|$ decreases (see figure \ref{fig:Q-Rem}). Thus we introduce the effective reconnection rate 

\begin{equation}
R^* \equiv -v_{yp} B_{xp}+v_{xp} B_{yp}
\end{equation}

which denotes the reconnection electric field normalized by $V_{A0} B_0$. 

The magnetic Reynolds number $R_{em}$ is defined as the ratio of the ambient Alfv\'{e}n speed to the magnetic diffusion speed. We should note that the shock-tube approximation adopted here is within a regime of ideal MHD, so we cannot treat the diffusion region itself in this scheme. Although one may think that we cannot argue the diffusion speed in this scheme at all, actually we can estimate it by a simple approximation as follows. In the vicinity of the reconnection point, the system is approximately in a steady state in the self-similarly expanding solution. This fact leads, from the Ampere's law, the electric field $E_z^* \equiv -v_{yp} B_{xp}+v_{xp} B_{yp}$ is approximately uniform around the diffusion region. The diffusion speed $v_{dif}^*$ must be balanced with the local inflow speed $v_{in}^*$ in the stationary state, and $v_{in}^*$ is estimated as $v_{in}^*=E_z^*/B_x^*=(-v_{yp} B_{xp}+v_{xp} B_{yp})/B_x^*$, where $B_x^*$ is the magnetic field just outside the diffusion region. We suppose $B_x^* \sim B_{xp}$ in order to approximately satisfy the magnetic flux conservation. 

Thus, we obtain the effective diffusion speed

\begin{equation}
v_{dif}^*=v_{in}^* \sim (-v_{yp} B_{xp}+v_{xp} B_{yp})/B_{xp} \ ,
\end{equation}

and define the effective magnetic Reynolds number

\begin{equation}
R_{em}^* \equiv 1/v_{dif}^* =B_{xp}/(-v_{yp} B_{xp}+v_{xp} B_{yp})\ .
\end{equation}

Note $V_{A0}=1$ in our normalization. Hence, we obtain

\begin{equation}
R^* = B_{xp}/R_{em}^* \ .
\end{equation}

It is very important to clarify the relation between $R^*$ and $R_{em}^*$. For this purpose, we treat $R_{em}^*$ as a control parameter instead of $v_{yp}$. The effective magnetic Reynolds number ($R_{em}^*$) dependence of the spontaneously expanding reconnection system in a free-space is investigated for the cases $\beta_0=0.01$ and $\beta_0=0.2$. 

We select the $y$-component $v_{yp}$ of the inflow velocity toward the slow shock as the artificially controllable variable for our numerical parameter survey. Starting at the solution with its spontaneous inhalation speed $v_{yp}=-v_{insp}(\beta_0)$ toward the slow shock and gradually decrease its absolute value with a finite decrement (see section \ref{sec:procedure}), and numerically solve the coupled equations listed in section \ref{sec:b-eqs} by the Newton-Raphson method for each value of $v_{yp}$. Thus we obtain a series of solutions for the outflow structure. Note that $R_{em}^*$ is a monotonically increasing function of $v_{yp}$ in our result (see figure \ref{fig:Q-Rem}), so we adopt $R_{em}^*$ as the control parameter in the following discussion. The result is as follows;\\

\noindent
1) Converging inflow\\
The converging speed $|v_{xp}|$ (note $v_{xp} < 0$ in our result which denotes a converging inflow) of the inflow drastically increases as the magnetic Reynolds number $R_{em}^*$ increases (see figure \ref{fig:Q-Rem}). The directions of the inflow velocity and the magnetic field tend to be parallel as the converging speed increases (see figure \ref{fig:R-Rem}). \\

\noindent
2) Magnetic Reynolds number dependence of reconnection rate\\
As $R_{em}^*$ increases, first the reconnection rate $R^*$ increases owing to the increase of $B_{xp}$ (see figure \ref{fig:Q-Rem}), and then $R^*$ decreases owing to that the directions of the inflow velocity and the magnetic field tend to be parallel (see figure \ref{fig:Q-Rem}). This causes a reduction of the reconnection electric field. Thus, the reconnection rate $R^*$ decreases as $R^* \propto 1/R_{em}^*$ because $B_{xp} \sim 1$ in the large $R_{em}^*$ region (see figure \ref{fig:Q-Rem}). \\

\section{Discussion}
\label{sec:discussion}

\subsection{Channel flow structure in the inflow}
\label{sec:channel}

For large magnetic Reynolds number (say, $R_{em}^* \geq 20$), a significantly converging inflow toward the slow shock arises. It can grow to be almost parallel to the magnetic field lines as $R_{em}^*$ increases. In such the case, the inflow toward the diffusion region must be very slow because the reconnection rate (note $R^* \equiv \mbox {\boldmath v}_p \times \mbox{\boldmath B}_p$) is very small, even though the Petschek-like slow shock still strongly intakes the plasma in order to satisfy the Rankine-Hugoniot condition. This means that the inflow forms a channel flow structure: relatively large speed converging inflow toward the slow shock and very slow inflow toward the diffusion region (see figure \ref{fig:channel}).

It is possible to observationally estimate the reconnection rate from the inflow Alfv\'{e}n Mach number (see Yokoyama et al. 2001, Isobe et al. 2005). If we can obtain the Alfv\'{e}n speed and the inflow speed perpendicular to the inflow magnetic field by observation (i.e., if we observe a motion of a bright filament-like structure toward the X-point, supposing it shows a field line motion, we can estimate the normal speed of the field line motion), we can estimate the reconnection rate as a good approximation. We must note, however, the inflow Alfv\'{e}n Mach number itself does not denote the reconnection rate if the inflow is inclined with respect to the magnetic field lines. Such the situation may occur when the flow is significantly converging (see section \ref{sec:result}). In such the case, the reconnection rate might be rather small even the inflow speed is very large. We must note that, in order to estimate the reconnection rate, we need not only the inflow Alfv\'{e}n Mach number, but also the direction of the magnetic field relative to the inflow velocity.

\subsection{ The role of the diffusion region} 

In a very short period (of the order of the Alfv\'{e}n transit time over the current sheet thickness) just after the onset of the reconnection, the evolution strongly depends on the resistivity model. Such information propagates by MHD waves. In the self-similar stage, since the effect of the resistivity model remains only in a thin layer in the vicinity of the FRWF, it does not influence the entire structure of our self-similar solution. When the induced inflow sufficiently evolves and Petschek-like slow shock forms, the slow shock begins to play major roles for spontaneous structure formation. The region around the diffusion region with a finite spatial scale seems also to be determined by the slow shock and tends to be stationary state as time proceeds. The resistive region plays only passive roles in this stage. These dynamics had been partially clarified by our numerical simulation (see Nitta et al. 2001). The discussion in this paper aims  to clarify the spontaneously determined structure lead by the shock-tube problem in the region described by ideal MHD. The following is just a conjecture in the present status, but I believe that it can be clarified in our near future works. 

In the self-similar stage, the size of the diffusion region (similar to the current sheet thickness or the ion Larmor radius) is infinitesimally small comparing with the entire system scale. Hence, the most of the inflow does not pass through the diffusion region, but pass through the Petschek-like slow shock to form the reconnection outflow. In such situation, it may be trivial that the entire structure should be determined not by the diffusion region, but by the Petschek-like slow shock. The only role of the diffusion region (but is not less meaningful) is to restrict the diffusion speed of the magnetic field lines. This effect of the diffusion region is clarified by this paper (see sections \ref{sec:result}  and \ref{sec:channel}). 

In my conjecture, the structure of the diffusion region in the self-similar stage is passively determined in a circumstance of the surrounding non-resistive region which is determined by the shock-tube problem involving the Petschek-like slow shock as is treated in this paper. In this case, the environmental circumstance which is determined by the shock-tube problem will play a role of a kind of boundary condition to determine the diffusion region. If we focus our attention into the diffusion region, this situation is very similar to the case of the ``driven reconnection'' which is also determined by the external boundary condition.

This intuition should be certified by solving the diffusion region under some matching condition with the environmental ideal MHD region. We must realize, however, that the structure of the diffusion region cannot be described by MHD but by some particle scheme, because the scale of the intended diffusion region is of the order of the ion Larmor radius. This is obviously out of the limit of MHD approximation. To inquire further into the matter would lead us into that specialized area of the particle reconnection, and such a digression would undoubtedly obscure the outline of our argument in the regime of the MHD reconnection.

\subsection{Petschek-type slow shock formation}
\label{sec:p-form}

In the shock-tube approximation adopted in this paper, we assume the steady formation of the Petschek-type figure-X shaped slow shock as a presupposition of the problem since we are interested in the fast reconnection regime. 

One may think this is curious because the entire system is strongly time-dependent. Though we cannot find any definite answer in the present status, the author would show the following conjecture. First, we must realize that the central region with a finite spatial scale around the diffusion region is in a steady state in the self-similar stage even the entire system is self-similarly expanding. In this case, the central region will be controlled by the surrounding stationary circumstance which is determined by the shock-tube problem (see the previous subsection). In naive intuition, we may expect a resultant stationary diffusion region and steady slow shock formation in the central region. The further study of this problem lies outside the scope of this paper as discussed in the previous subsection.  

We must note that the assumption of steady slow shock formation might not be appropriate for the case with a very small reconnection rate. The converging speed $|v_{xp}|$ drastically increases as the magnetic Reynolds number $R_{em}^*$ increases (see figure \ref{fig:Q-Rem}), and it exceeds unity at $R_{em}^* \sim 50$. This is very curious. As long as the inflow is spontaneously induced by the reconnection system, its speed may not exceed unity for the case of low $\beta_0$ because the Alfv\'{e}n speed is the maximum speed when magnetic energy is completely converted to the bulk kinetic energy. This curious result might owing to our presupposition that the Petschek type slow shock always forms: In order to keep to form the slow shock, the Rankine-Hugoniot condition requires unrealistic very high speed inflow. 

In actual reconnection, if the resistivity is very small and hence the magnetic Reynolds number becomes large (say, $R_{em}^* > 50$), the Petschek type slow shock may not form. Thus, reconnection may change its type to another one with no slow shock (e.g., the Sweet-Parker type; Sweet [1958], Parker [1963]) or X-O-X type one (see subsection \ref{sec:xox}). For the Sweet-Parker type of reconnection, we cannot adopt a shock-tube approximation, and it should be beyond the scope of this paper.

\subsection{Coupling between microscopic and macroscopic processes}
\label{sec:scale-coupling}

The size of the diffusion region seems to be extremely small if we estimate it by of the order of ion Larmor radius ($\sim 10^0$[m] for typical solar corona). From a naive intuition, it is very curious that such a very small diffusion region governs the huge entire reconnection system (e.g., the radius of magnetic flux tube, say $\sim 10^7$[m] for typical solar flares). Shibata \& Tanuma (2001) proposed a new idea of ``fractal reconnection'' that effective diffusion region has a fractal structure in which total scale of the diffusion region is not so small, and is composed of spatially self-similar small structures. Our self-similar evolution model, however, gives an alternative picture about this problem as follows. 

The self-similar reconnection model grows in a very wide dynamic range of its spatial scale. It may be triggered when the system scale is of the order of ion Larmor radius which is the scale of the current sheet thickness to drive an anomalous resistivity. The self-similar solution can approximate the actual phenomenon when the reconnection system scale is much larger than ion Larmor radius (say, $> 10^2$[m]). This self-similar solution can describe the system until the expanding system scale reaches an environmental proper scale which is imposed by external circumstances. Hence, we must note that the self-similar reconnection model is describing the coupling process between the microscopic scale and the macroscopic scale.  

Needless to say, the major energy release is determined by the largest structure of the reconnection system. There are two different points of view about how is the largest structure determined. In the most of previous works, it is treated to be governed by external circumstances through the boundary condition (the externally ``driven'' reconnection, see Vasyliunas 1975, Priest \& Forbes 1986, Sato \& Hayashi 1979, etc.). In this scheme, the external circumstances directly determine the large scale structure independent of the microscopic process of the diffusion region. In another scheme, the large scale structure is determined by intrinsic causes which properly included in the system itself, so we call it the ``spontaneous'' reconnection. Ugai \& Tsuda (1977) first showed a spontaneous model: the resistivity model denoting microscopic processes of the anomalous resistivity determines the entire system. Our self-similar evolution model is an advanced spontaneous reconnection model extended to the time-dependent reconnection which may be actual in astrophysical applications. 

The evolution just after the onset of reconnection will depend on detail properties of the resistivity model, i.e., size and shape of the resistive region, value and distribution of the resistivity inside the diffusion region, etc. Anyway, once reconnection starts, its size expands in the fast-mode rarefaction wave (FRW) propagation speed. When the system scale (i.e., the scale of the wave front FRWF of FRW) becomes much larger than the initial scale (the initial current sheet thickness $\sim$ ion Larmor radius), the system tends to be described by our self-similar solution. After this stage, the reconnection will proceed unlimitedly in a self-similar way. The self-similar solution depends only on two parameters (the plasma-$\beta$ value $\beta_0$ at the asymptotic region and the effective magnetic Reynolds number $R_{em}^*$), and is independent of other properties. For example, effects of above mentioned detailed properties of the resistivity model is contracted within a very narrow region in the vicinity of FRWF. As the self-similar expansion proceeds, the influence of detailed properties is quickly diminished. Because the self-similar solution well approximates the system as long as its scale is much larger than the initial current sheet thickness ($\sim 10^0$[m]), our self-similar solution is valid after the system scale becomes, say $10^2$[m]. This is very small comparing with the final system scale ($\sim 10^7$[m]). After that stage, the structure does not change no more but simply expands self-similarly. Thus, we should realize that the self-similar solution which is determined when the system scale is very small ($\sim 10^2$[m]) governs much larger structure in a later stage by the manner of self-similar expansion. Even the system scale reaches some external scale, the properties of the self-similar solution hold for a very long period ($\sim$ several hundred times the fast-mode transit time, see section 4.5 of paper 1).

\subsection{Steady vs. time-dependent and driven vs. spontaneous}
\label{sec:driven-spon}

In ``time-dependent'' reconnection models starting from an equilibrium state, a plasmoid forms around the contact discontinuity between the reconnection jet and the initial current sheet plasma. This plasmoid is a result of the interaction between these two different plasmas. In our ``spontaneous'' model in a free-space, the entire system is self-consistently determined from this interaction (see Nitta et al. 2002 and Nitta 2004). Thus, we do not need any external boundary condition in order to determine the resultant reconnection system.  

On the other hand, in ``steady'' models (see Vasyliunas 1975 or Priest and Forbes 1986), the information of this interaction is completely lost. Thus, we need to impose the external boundary condition in order to determine the reconnection system. This is so-called a ``driven'' type reconnection. A similar discussion about the difference between steady and time-dependent reconnection is in Ugai and Zheng (2005). 

Needless to say, the answer to the question which type is relevant depends on the circumstance of the problem. The author had been discussed the relation between the spontaneous model and the driven model in paper 3 (see section 5.6) in detail. I here emphasize again that these two different models are not opposite subjects, but models for different phases of the entire evolutionary process of the astrophysical reconnection. In my opinion, the ``driven phase'' follows the ``spontaneous phase'' in usual astrophysical problems as follows. 

A large scale plasma convection (e.g., convection of flux tubes in the solar plasma) will form a current sheet system, and stores an enormous magnetic energy into the current sheet system. In usual cases of convection with the speed at very small Alfv\'{ e}n Mach number, it will need a very long time comparing with the Alfv\'{e}n transit timescale for the energy storage (e.g., in typical solar flares, it needs several days $\sim 10^5$ [s]). As the result of current sheet thinning, once a well-localized anomalous resistivity is switched on in the current sheet, it triggers the onset of a fast reconnection. The evolution quickly moves to the self-similar expansion. The reconnection system expands at the speed of the FRW propagation which is estimated as the ambient Alfv\'{e}n speed (much faster than the convection speed). Hence, the evolution just after the onset is ``spontaneous'' one, because the timescale of the expansion is much smaller ($\sim 10^2$ [s] for typical solar flares) than the timescale imposed by the external circumstance ($\sim 10^5$ [s]). When the FRWF reaches the ambient circumstance, the evolution will start to be influenced by the external boundary condition, but the central energy conversion region holds as the spontaneous state during almost one hundred times the Alfv\'{e}n transit timescale (see section 4.5 of Nitta et al. 2001). After the central region is altered by the external boundary condition, the system moves to the ``driven'' phase at last.

\subsection{Possibility of another new type of solution}
\label{sec:xox}

Many people believe that the Petschek type fast reconnection can realize under the existence of the locally enhanced anomalous resistivity (Biskamp 1986, Scholer 1989, Yokoyama \& Shitaba 1994). We have not been able to identify the mechanism of the anomalous resistivity yet, but some kind of current driven microscopic instability should be the origin of the anomalous resistivity. In naive picture, such current driven anomalous resistivity may arise at the locus where the current sheet is strongly compressed. 

In this paper, we considered only the case that the diffusion region is fixed around the origin of the coordinate. If we apart from this presumable assumption, we may realize another possibility. From the discussion of the channel flow structure (see subsection \ref{sec:channel}), we can speculate the following new type of solutions. In the channel flow, the inflow speed toward the slow shock is much larger than it toward the diffusion region. This non-uniform inflow leads a non-uniform compression of the field reversal region. Since the compression outside the diffusion region is much stronger owing to a strong inhalation by the slow shock, the reconnection point may move outside the present diffusion region. A bipolar reconnection outflow will be ejected from the new X-point. One outflow towards the direction apart from the previous diffusion region, another towards the previous diffusion region. Thus, an O-point will form around the original diffusion region. Consequently, a typical X-O-X structure will form. We must note that this X-O-X structure is not a result of the tearing-mode instability but of the non-uniform inflow speed lead by the Petschek-like slow shock. 

The existence of this new solution shows a possibility of the bifurcation of the solution. If the magnetic Reynolds number $R_{ em}$ is small enough (say, $< 20$), the Petschek-like X-point structure will be stable because no channel flow structure forms. When $R_{ em}$ increases and exceeds a critical value ($\sim \mbox{several}  \times 10$), a remarkable channel flow structure will take place, and the solution will bifurcate into two different types of solutions: the single X-point solution and the X-O-X solution. 

A similar discussion about non-uniqueness of the reconnection solutions was shown in Hameiri (1979). He applied his model to flare phenomena as abrupt jumps to different branches of the solution. The answer to the question, in our case, which solution is stable will depend on the resistivity model (artificially fixed type around the origin or a current driven type) and the ``history'' of the time variation of the magnetic Reynolds number. It may result in a hysteresis behavior. An indication of the transition of the solution to the X-O-X structure is appeared in our preliminary numerical simulation. The details will be discussed in our following papers. 

We must realize, however, that we cannot completely avoid these kinds of ambiguity arising from the non-uniqueness of the resistivity model as long as our study is restricted within the regime of MHDs because the resistivity cannot be determined from the MHD equations. Unfortunately, we have not had any convincing microscopic model of anomalous resistivity. I believe that these facts do not reduce the importance to study the MHD reconnection; on the contrary, we must study MHD reconnection more as an elementary process in astrophysical plasma before applying to complicated global problems because we have not established complete astrophysical MHD reconnection model yet.

\vspace{1cm}

The author thanks to Atsuhiro Nishida (SOKENDAI), Kazunari Shibata (Kyoto Univ.), Takaaki Yokoyama (Tokyo Univ.), Takahiro Kudoh (National Astronomical Observatory of Japan) and Syuniti Tanuma (Kyoto Univ.) for fruitful scientific discussion and comments. Also thanks to Tetsuyuki Yukawa (SOKENDAI) for the comments on numerical procedure. Improvement of English usage is owing to the advice from Naoko Kato (SOKENDAI).

\appendix
\section{Equations for structure of the reconnection outflow}

The structure of the reconnection outflow is determined by the following equations. \\

We assume that the region between the asymptotic region and the pre-shock region is filled with non-resistive plasma. Hence, the magnetic flux is frozen into the induced inflow. We also assume a polytropic variation in the induced inflow because there is no violent process in the fast-mode rarefaction. 
Thus, we impose\\

\noindent
Frozen-in condition [eq1]
\begin{equation}
\frac{\sqrt{B_0^2-B_0^2 v_{xp}^2/(v_{xp}^2+v_{yp}^2)}}{\rho_0}=\frac{\sqrt{B_{xp}^2+B_{yp}^2-(B_{xp} v_{xp}+B_{yp}v_{yp})^2/(v_{xp}^2+v_{yp}^2)}}{\rho_p} \label{eq:frozen-in}
\end{equation}

and \\

\noindent
Polytropic relation [eq2]
\begin{equation}
P_0 \rho_0^{-\gamma}=P_p \rho_p^{-\gamma} \ .
\end{equation}
where $\gamma$ is the specific heat ratio. We assume $\gamma=5/3$ (monoatomic ideal gas).

There are several discontinuities in the reconnection outflow, i.e., X-shaped slow shock, reverse fast shock, contact discontinuity, and forward V-shaped slow shock (see figure \ref{fig:sch}). We set jump conditions for both sides of each discontinuity:\\

\noindent
X-shaped slow shock Rankine-Hugoniot (R-H) jump conditions\\

\noindent
Pressure-jump [eq3]
\begin{eqnarray}
\frac{P_1}{P_p}=&1+\frac{\gamma}{c_{sp}^2}(-\sin \theta v_{xp} + \cos \theta v_{yp})^2 (X-1) \times
\{
\frac{1}{X}-(\cos \theta B_{xp}+\sin \theta B_{yp})^2/2 \nonumber \\
&\times [-2 V_{Apx}^2 X+(-\sin \theta v_{xp} + \cos \theta v_{yp})^2 (X+1)]/[((-\sin \theta v_{xp} + \cos \theta v_{yp})^2-X V_{Apx}^2)^2 \mu \rho_p]
\}
\end{eqnarray}

where
$$c_{sp}=\sqrt{\gamma P_p/\rho_p} \ ,$$ 
$$V_{Apx}=\sqrt{(-\sin \theta B_{xp}+\cos \theta B_{yp})^2/(\mu \rho_p)} \ ,$$
$\mu$ is the magnetic permeability of vacuum and $X$ is the compression ratio. \\

\noindent
Density-jump [eq4]
\begin{equation}
\frac{\rho_1}{\rho_p}=X
\end{equation}

\noindent
Velocity (parallel) jump [eq5]
\begin{equation}
\frac{\cos \theta v_1-v_0}{\sin \theta v_{yp}-v_0}=\frac{(-\sin \theta v_{xp} + \cos \theta v_{yp})^2-V_{Apx}^2}{(-\sin \theta v_{xp} + \cos \theta v_{yp}^2-X V_{Apx}^2)}
\end{equation}

where $v_0=(v_{yp} B_{xp}-B_{yp} v_{xp})/(B_{xp} \sin \theta-B_{yp} \cos \theta)$ is the shift speed of the de Hoffmann-Teller coordinate.

\noindent
Velocity (perpendicular) jump [eq6]
\begin{equation}
\frac{-\sin \theta v_1}{-\sin \theta v_{xp} + \cos \theta y_{yp}}=\frac{1}{X}
\end{equation}

\noindent
Magnetic field (parallel) jump [eq7]
\begin{equation}
\frac{\sin \theta B_1}{\cos \theta B_{xp}+\sin \theta B_{yp}}=\frac{[(-\sin \theta v_{xp} + \cos \theta v_{yp})^2-V_{Apx}^2]X}{(-\sin \theta v_{xp} + \cos \theta v_{yp})^2-X V_{Apx}^2}
\end{equation}

\noindent
Magnetic field (perpendicular) jump [eq8]
\begin{equation}
\frac{\cos \theta B_1}{-\sin \theta B_{xp}+\cos \theta B_{yp}}=1
\end{equation}

The compression ratio $X$ is defined by the following equation (3rd order algebraic equation for $X$),
\begin{eqnarray}
&\{
(-\sin \theta B_{xp}+\cos \theta B_{yp})^2 [(\cos \theta B_{xp}+\sin \theta B_{yp})^2 (\gamma - 1) \rho_p (-\sin \theta v_{xp} + \cos \theta v_{yp})^2 \nonumber \\
&+(-\sin \theta B_{yp}+\cos \theta B_{yp})^2 (2 \gamma P_p-\rho_p (-\sin \theta v_{xp} + \cos \theta v_{yp})^2+\gamma \rho_p (-\sin \theta v_{xp} + \cos \theta v_{yp})^2)]
\} X^3 \nonumber \\
&+
\{
-\rho_p (-\sin \theta v_{xp} + \cos \theta v_{yp})^2 
[
(-\sin \theta B_{xp}+\cos \theta B_{yp})^4 (\gamma-1) \nonumber \\
&+
(\cos \theta B_{xp}+\sin \theta B_{yp})^2 (\gamma-2) \mu \rho_p (-\sin \theta v_{xp} + \cos \theta v_{yp})^2 \nonumber \\
&+
(-\sin \theta B_{xp}+\cos \theta B_{yp})^2
(
(\cos \theta B_{xp}+\sin \theta B_{yp})^2 (\gamma+1) \nonumber \\
&+
2 \mu (2 \gamma P_p-\rho_p (-\sin \theta v_{xp} + \cos \theta) v_{yp})^2+\gamma \rho_p (-\sin \theta v_{xp} + \cos \theta v_{yp})^2
)
]
\} X^2 \nonumber \\
&+
\{
\mu \rho_p (-\sin \theta v_{xp} + \cos \theta v_{yp}^4
[
(\cos \theta B_{xp}+\sin \theta B_{yp})^2 \gamma+
2(-\sin \theta B_{xp}+\cos \theta B_{yp})^2 (\gamma+1)+
\mu \nonumber \\
&(
2 \gamma P_p-\rho_p (-\sin \theta v_{xp} + \cos \theta v_{yp})^2+\gamma \rho_p (-\sin \theta v_{xp} + \cos \theta v_{yp})^2
)
]
\} X \nonumber \\
&+
\{
-(\gamma+1) \mu^2 \rho_p^3 (-\sin \theta v_{xp} + \cos \theta v_{yp})^6
\}=0 \ . \nonumber
\end{eqnarray} 
This equation has three roots. We must choose a real positive root larger than unity.

\noindent
Reverse-fast shock R-H conditions\\

\noindent
Pressure-jump [eq9]
\begin{equation}
\frac{P_2}{P_1}=\zeta_f
\end{equation}

where 
$$\zeta_f=\gamma M_{1f}^2 (1-1/\xi_f)+(1-\xi_f^2)/\beta_1+1$$ 
with 
$$\xi_f=(-l + \sqrt{l^2 + 2/\beta_1 (2 - \gamma) (\gamma + 1) \gamma M_{1f}^2})/(2/\beta_1 (2 - \gamma)) \ ,$$
$$l = \gamma (1/\beta_1 + 1) + (\gamma - 1) \gamma M_{1f}^2/2 \ ,$$
$$\beta_1 = P_1/(B_1^2/(2 \mu)) \ ,$$
$$M_{1f} = (v_1 - x_f V_A)/c_f \ ,$$
$$c_f = \sqrt{\gamma P_1/\rho_1}$$
and 
$$V_A = B_0/\sqrt{\mu \rho_0}\ .$$

\noindent
Density-jump [eq10]
\begin{equation}
\frac{\rho_2}{\rho_1}=\xi_f
\end{equation}

\noindent
Velocity jump [eq11]
\begin{equation}
\frac{v_1-x_f}{v_2-x_f}=\xi_f
\end{equation}

\noindent
Magnetic field jump [eq12]
\begin{equation}
\frac{B_2}{B_1}=\xi_f
\end{equation}

We must impose local magnetic flux conservation on both sides of region p and region 1. \\

\noindent
Magnetic flux conservation at X-point [eq13]
\begin{equation}
-v{xp} B_{yp} + v_{yp} B_{xp}+v_1 B_1=0
\end{equation}

\noindent
Force balance at contact discontinuity [eq14]
\begin{equation}
P_2+\frac{B_2^2}{2 \mu}=P_3+\frac{B_{x3}^2+B_{y3}^2}{2 \mu}
\end{equation}

\noindent
V-shaped forward slow shock R-H conditions\\

\noindent
Pressure-jump [eq15]
\begin{equation}
P_3=P_0 \left\{
1+\frac{\gamma}{c_{s0}^2} (x_s \sin \phi)^2 (X_h-1)
\times \left[
\frac{1}{X_h}-\frac{(B_0 \cos \phi)^2}{2} \frac{-2 V_{A0x}^2 X_h+(x_s \sin \phi)^2 (X_h+1)}{[(x_s \sin \phi)^2-X_h V_{A0x}^2] \mu \rho_0}
\right]
\right\}
\end{equation}
where
$c_{s0}=\sqrt{\gamma P_0/\rho_0}$, $V_{A0x}=\sqrt{(-\sin \theta B_0)^2/(\mu \rho_0)}$ and $X_h$ is the compression ratio. \\

\noindent
Density-jump [eq16]
\begin{equation}
\frac{\rho_3}{\rho_0}=X_h
\end{equation}

\noindent
Velocity (parallel) jump [eq17]
\begin{equation}
\frac{\cos \phi (v_{x3}-x_s V_{A0})+\sin \phi v_{y3}}{-V_{A0} x_s \cos \phi}
=\frac{v_{m0x}^2-V_{A0x}^2}{v_{m0x}^2-X_h V_{A0x}^2}
\end{equation}
where $v_{m0x}=x_s \sin \phi$ and $V_{A0}=B_0/\sqrt{\mu \rho_0}$.

\noindent
Velocity (perpendicular) jump [eq18]
\begin{equation}
\frac{-\sin\phi (v_{x3}-x_s V_{A0})+\cos \phi v_{y3}}{V_{A0} x_s \sin \phi}=\frac{1}{X_h}
\end{equation}

\noindent
Magnetic field (parallel) jump [eq19]
\begin{equation}
\frac{\cos \phi B_{x3}+\sin \phi B_{y3}}{B_0 \cos \phi}=\frac{(v_{m0x}^2-V_{A0x}^2) X_h}{v_{m0x}^2-X_h V_{A0x}^2}
\end{equation}

\noindent
Magnetic field (perpendicular) jump [eq20]
\begin{equation}
\frac{-\sin \phi B_{x3}+\cos \phi B_{y3}}{-B_0 \sin \phi}=1
\end{equation}

The compression ratio $X_h$ is a solution of the following third order algebraic equation:
\begin{eqnarray}
&\{
(B_0 \sin \phi)^2 
[
(B_0 \sin \phi)^2 (\gamma-1) \rho_0 (x_s \sin \phi)^2+
(B_0 \sin \phi)^2 
(
2 \gamma P_0-\rho_0 (x_s \sin \phi)^2+\gamma \rho_0 (x_s \sin \phi)^2
)
]
\} X_h^3 \nonumber \\
&+
\{
-\rho_0 (x_s \sin \phi)^2
[
(B_0 \sin \phi)^4 (\gamma+1)+(B_0 \cos \phi)^2 (\gamma-2) \mu \rho_0 (x_s \sin \phi)^2 \nonumber \\
&+(B_0 \sin \phi)^2 
(
(B_0 \cos \phi)^2 (\gamma+1)+(2 \mu (2 \gamma P_0-\rho_0 (x_s \sin \phi)^2+\gamma \rho_0 (x_s \sin \phi)^2))
)
]
\} X_h^2 \nonumber \\
&+
\{
\mu \rho_0^2 (x_s \sin \phi)^4 
(
(B_0 \cos \phi)^2 \gamma + 2(B_0 \sin \phi)^2 (\gamma+1)+
\mu (2 \gamma P_0 - \rho_0 (x_s \sin \phi)^2+\gamma \rho_0 (x_s \sin \phi)^2)
)
\} X_h \nonumber \\
&+
\{
-(\gamma+1) \mu^2 \rho_0^3 (x_s \sin \phi)^6
\}=0\ . \nonumber
\end{eqnarray}

We must choose a real positive root larger than unity. 

The tip of the reconnection outflow touches the FRWF (see figure \ref{fig:sch}), hence $A_1'=0$ at this point. This condition reduces to the following equation. 

\noindent
Boundary Condition at the tip of the outflow [eq21]
\begin{equation}
-B_{y3} (1-x_c) + B_{x3} [(1-x_s) \tan \phi - x_c \tan \theta]-B_0 (1-x_s) \tan \phi =0
\end{equation}

From magnetic flux conservation, the injected magnetic flux must be redistributed in the reconnection jet. This leads to the following equation. 

\noindent
Magnetic flux conservation [eq22]
\begin{equation}
-B_{yp} v_{xp}+v_{yp} B_{xp}+[x_f B_1+ (x_c-x_f) B_2]=0
\end{equation}

We assume the following trivial relations:\\

\noindent
Definition from the contact discontinuity
$$
x_c=v_2=v_{x3} 
$$

We can solve [eq2] for $P_p$, [eq8] for $B_1$, [eq9] for $P_2$, [eq10] for $\rho_2$, [eq12] for $B_2$, [eq13] for $v_{yp}$, [eq15] for $P_3$, [eq16] for $\rho_3$, [eq17] and [eq18] for $v_{x3}$ and $v_{y3}$, [eq19] and [eq20] for $B_{x3}$ and $B_{y3}$ by hand, then substitute them into other nine equations ([eq1], [eq3], [eq4], [eq5], [eq6], [eq7], [eq11], [eq14], [eq21] and [eq22]) for the following unknowns: $\rho_p$, $v_{xp}$, $B_{xp}$, $B_{yp}$, $\theta$, $P_1$, $\rho_1$, $x_f$, $\phi$ and $x_s$. The only parameter included in this problem is the plasma-$\beta$ value at the asymptotic region. By using a Newton-Raphson routine, with an initial guess for these unknowns, we obtain converged solutions. The procedure to obtain the series of converged solutions is discussed in section \ref{sec:b-eqs} in detail.

As $|v_{yp}|$ decreases, the strength of reverse fast shock reduces, and then the pressure jump $\zeta_f$ and density jump $\xi_f$ simultaneously become unity at the critical value of $|v_{yp}|$. At that point, the reverse fast shock vanishes. We convert the coupled equations to@another set compatible to the situation with no fast shock, i.e., several equations reform to the following equations: 

\noindent
[eq9] is replaed to [eq9']\\
$$P_2=P_1 \ ,$$ 

\noindent
[eq10] is replaed to [eq10']\\
$$\rho_2=\rho_1 \ ,$$ 

\noindent
[eq11] is replaed to [eq11']\\
$$v_1=v_2 \ ,$$

\noindent
[eq12] is replaed to [eq12']\\
$$B_1=B_2 \ ,$$ 

and [eq22] becomes to be equivalent to [eq13] and removed from the set of coupled equations. The number of coupled equations reduces to 21. This is consistent with that the locus $x_f$ of the fast shock is no longer included in the set of unknowns and total number of unknowns becomes 21.

\appendix

\figcaption[]{Spontaneous inhalation speed toward the X-slow shock. The $y-$component $v_{yp}$ of the velocity at the region p versus the ambient plasma-$\beta$ for the case $v_{xp}=0$ (non-converging inflow) is shown. Similar to this result of $v_{yp}$, we can find any unknown variable as a function of $\beta_0$. These solutions for arbitrary $\beta_0$ can be treated as initial guess for the procedure in subsection \ref{sec:procedure}. This figure is almost the same to figure 6 of Nitta 2004, but the following points are different: much higher converging precision, precise treatment for junction between fast shock regime and no-fast shock regime.  
\label{fig:sp-in}
}

\figcaption[]{$\beta_0$-dependence of the reconnection rate $R^*$. The reconnection rate is almost constant ($\sim 0.05$) as a function of $\beta_0$ (plasma-$\beta$ at the asymptotic region far outside the current sheet). The obtained value of the reconnection rate is consistent with previous theoretical and numerical works, but we must note that this value is spontaneously determined by the reconnection system itself in a frame work of ideal (non-resistive) MHD. This figure is almost the same to figure 2 of Nitta 2004, but the following points are different: much higher converging precision, precise treatment for junction between fast shock regime and no-fast shock regime.  
\label{fig:R-beta0}
}

\figcaption[]{Schematic figure of the reconnection outflow. Several discontinuities form in the outflow. We consider a quasi-one dimensional shock-tube problem along the $x-$ and $y-$axes. 
\label{fig:sch}
}

\figcaption[]{Velocity and magnetic field at the region p versus magnetic Reynolds number $R_{em}^*$. This is the case for $\beta_0=0.01$. These quantities are essential to determine the reconnection rate $R^*$. As $R_{em}^*$ increases, first, the transverse component $B_{xp}$ of the magnetic field quickly increases. Then, $v_{xp}$ drastically developed. The negative value of $v_{xp}$ denotes a converging inflow toward the X-slow shock. 
\label{fig:Q-Rem}
}

\figcaption[]{Direction of velocity and magnetic field at the region p. This is the case for $\beta_0=0.01$. The circular dot and square denote $v_{yp}/v_{xp}$ (tangent of the inclination angle of the inflow velocity) and $B_{yp}/B_{xp}$ (tangent of the inclination angle of the inflow magnetic field), respectively. This graph  clearly shows that the inflow velocity $\mbox{\boldmath v}_p$ and the inflow magnetic field $\mbox{\boldmath B}_p$ tend to be parallel to each other when the magnetic Reynolds number $R_{em}^*$ increases. This causes the decrement of the reconnection rate $R^* \equiv \mbox{\boldmath v}_p \times \mbox{\boldmath B}_p$.  
\label{fig:direction-vB-Rem}
}

\figcaption[]{Magnetic Reynolds number dependence of the reconnection rate $R^*$. This is the case for $\beta_0=0.01$. As $R_{em}^*$ increases from the case $v_{xp}=0$ (spontaneous inhalation), first $R^*$ increases because $B_{xp}$ increases (see figure \ref{fig:Q-Rem}), then $R^*$ decreases as $\propto 1/R_{em}^*$. 
\label{fig:R-Rem}
}

\figcaption[]{Schematic picture of channel flow structure. DR: diffusion region, SS: Petschek-like slow shock, SSL: separatrix stream line which divides the stream lines reaching the Petschek-like slow shock and the stream lines reaching DR. Solid lines show the magnetic field lines. Arrows show the flow velocity. This is the case of significantly converging inflow. The right side of SSL is filled with uniform high speed converging inflow, while the left side of SSL is filled with very slow speed inflow. 
\label{fig:channel}
}

\end{document}